\documentclass[twocolumn,twoside]{revtex4}
\usepackage{graphicx}
\usepackage{fancyhdr}
\usepackage{hyperref}
\usepackage{amsmath,amssymb}

\usepackage{longtable} 
\usepackage{tabularx}
\usepackage{multirow}
\usepackage{array}
\newcolumntype{C}[1]{>{\centering\arraybackslash}p{#1}}
\newcolumntype{L}[1]{>{\raggedright\arraybackslash}p{#1}}
\newcolumntype{R}[1]{>{\raggedleft\arraybackslash}p{#1}}

\usepackage{xcolor}

\begin{document}

\title{Theoretical Review of Rare B Decays}

%

\author{F.~Mahmoudi}

\affiliation{Universit\'e de Lyon, Universit\'e Claude Bernard Lyon 1, CNRS/IN2P3,\\
Institut de Physique des 2 Infinis de Lyon, UMR 5822, F-69622, Villeurbanne, France}

\affiliation{Theoretical Physics Department, CERN, CH-1211 Geneva 23, Switzerland}

\begin{abstract}
We present an overview of the recent neutral current $B$ decays focusing on the deviations with respect to the Standard Model predictions that are observed in $b \to s \ell^+\ell^-$ transitions, and discuss their implications for new physics scenarios in a model-independent way by means of global statistical fits. The prospects for future discoveries in the individual channels, in particular using the theoretically clean observables are also addressed. 
\end{abstract}

\maketitle

\thispagestyle{fancy}


\section{Introduction}
Rare $B$ decays are important probes for physics beyond the Standard Model (SM) as they are loop-suppressed in the SM and are therefore sensitive to new physics (NP) parameters. As sensitive flavour observables, they are connected to fundamental questions in particle physics and in particular play a key role in understanding the underlying pattern of the SM. Furthermore, they can provide guidance for NP model-building.

In the recent years, there have been several experimental measurements showing deviations with respect to the SM predictions, that are commonly referred to as flavour anomalies. The angular observable $P_5^\prime$ of the $B \to K^* \mu^+ \mu^-$ decay~\cite{Descotes-Genon:2012isb} was the first one measured by LHCb collaboration in 2013~\cite{LHCb:2013ghj} presenting more than 3$\sigma$ discrepancy with the SM. Updated measurements by LHCb have persistently confirmed this tension, which can be explained with short distance new physics contributions to this decay~\cite{LHCb:2015svh,LHCb:2020lmf}. 
The overall $B \to K^* \mu^+ \mu^-$ angular observables are in agreement (see e.g.~\cite{Hurth:2020ehu}) and the tension is also supported by the recent angular analysis of $B^+ \to K^{*+} \mu^+ \mu^-$~\cite{LHCb:2020gog}. Another decay indicating tension with the SM is $B_s \to \phi \mu^+ \mu^-$~\cite{LHCb:2015wdu,LHCb:2021xxq,LHCb:2021zwz}, in particular its branching ratio is measured to be below SM prediction.
This trend is observed in several other $b\to s \ell^+ \ell^-$ branching ratios such as $B \to K \mu^+ \mu^-$~\cite{LHCb:2014cxe} and $\Lambda_b \to \Lambda \mu^+ \mu^-$~\cite{LHCb:2015tgy}. 
However, the branching ratios are dependent on the local form factors and suffer from large theoretical uncertainties~\cite{Grozin:1996pq,Kruger:1999xa,Beneke:2001at,Asatryan:2001zw,Grinstein:2004vb,Beneke:2004dp,Kruger:2005ep,Khodjamirian:2010vf,Beylich:2011aq,Nishikawa:2011qk,Khodjamirian:2012rm,Hambrock:2013zya,Lyon:2014hpa}. The angular observables on the other hand, while having a reduced sensitivity to the form factor uncertainties~\cite{Egede:2008uy,Egede:2010zc}, receive still non-local contributions that are not fully under control in QCD factorisation~\cite{Hurth:2016fbr}. Therefore, the significance of the deviations depend on the estimated size of the non-local effects. Recent theoretical progress has been achieved for a better control of these effects in Refs.~\cite{Bobeth:2017vxj,Braun:2017liq,Gubernari:2020eft,Gubernari:2022hxn}.

Furthermore, a set of observables is defined in order to check lepton flavour universality violation (LFUV) in $b \to s \ell^+\ell^-$ decays as~\cite{Hiller:2003js}
\begin{equation}
R_H = (B\to H \mu^+\mu^-)/(B\to H e^+e^-)\, ,
\end{equation}
with $H=K^+,K^*,\phi$, etc.
Such ratios, contrary to the observables mentioned above, are very clean and precisely known in the SM (see Ref.~\cite{Isidori:2022bzw} for a recent study of the QED corrections for these observables).
Deviations from the SM predictions are seen in the LFUV ratios for $R_K$~\cite{LHCb:2014vgu,LHCb:2019hip,LHCb:2021trn} and $R_{K^*}$~\cite{LHCb:2017avl} that are measured to be below the SM predictions. Similarly, the recent measurements of $R_{K_S^0}$ and $R_{K^{*+}}$~\cite{LHCb:2021lvy} although within 2$\sigma$ of the SM predictions, show the same trend with the central values below the SM predictions.

Interestingly, modest signs of LFUV are also observed in flavour changing neutral current processes in the Kaon sector~\cite{DAmbrosio:2022kvb}.

Another theoretically clean observable with an uncertainty of less than 5\% is the branching ratio of $B_s \to \mu^+ \mu^-$ which has been measured by several experiments. Here we consider the 
combination of ATLAS~\cite{ATLAS:2018cur}, CMS~\cite{CMS:2019bbr} and LHCb~\cite{LHCb:2021awg,LHCb:2021vsc} as given in Ref.~\cite{Hurth:2021nsi}.

While the significance of the reported anomalies taken individually is around $\sim2-3\sigma$, when considered collectively they can find a common NP explanation with a much larger significance in a global analysis~\cite{Hurth:2021nsi,Alguero:2021anc,Altmannshofer:2021qrr,Ciuchini:2020gvn,Geng:2021nhg,London:2021lfn}.

\section{Theoretical setup and uncertainties}
The description of $b \to s  \ell^+ \ell^-$ transitions is based on the effective Hamiltonian

\begin{eqnarray}
 {\cal H}_{\rm eff} &=& -\frac{4G_F}{\sqrt{2}}V_{tb}V_{ts}^* \Big\{ \sum_{i=1,\ldots,6,8}C_i\; O_i \\
 && + \sum_{i=7,9,10,Q_1,Q_2,T}(C_i\; O_i + C^\prime_i\; O^\prime_i)\,\nonumber \Big\}\, ,
\end{eqnarray}
where $G_F$, $V_{tb}$ and $V_{ts}$ are the Fermi constant and two CKM matrix elements, respectively, and the $O_i$ are local operators coming each with an associated Wilson coefficient $C_i$ which is calculable perturbatively for a particular high-energy physics model.

The semileptonic part of the Hamiltonian (second term in the left hand side) accounts for the dominant contribution and can be factorised into a leptonic and a hadronic piece. The latter can be described by seven independent form factors $\tilde{S}, \tilde{V}_\lambda, \tilde{T}_\lambda$, with helicities $\lambda=\pm1,0$. 
The hadronic part of the Hamiltonian (first term in the left hand side) has a subleading contribution to $B\to K^* \mu^+ \mu^-$ from a virtual photon decaying into a lepton pair. This leads to non-factorisable contributions and appears in the vectorial helicity amplitudes
\begin{eqnarray}\label{eq:HV}     
 H_V(\lambda) &=& -i\, N^\prime \Big\{ C_9^{\rm eff} \tilde{V}_{\lambda} - C_{9}'  \tilde{V}_{-\lambda}\\
& +& \frac{m_B^2}{q^2} \Big[\frac{2\,\hat m_b}{m_B} (C_{7}^{\rm eff} \tilde{T}_{\lambda} - C_{7}'  \tilde{T}_{-\lambda})
      - 16 \pi^2 {\cal N}_\lambda \Big] \Big\}\, , \nonumber
\end{eqnarray}
where the factorisable piece is described as the effective part of  $C_9^{\rm eff}\left(\equiv C_9+Y(q^2)\right)$ and the non-factorisable piece is encoded in ${\cal N}_\lambda(q^2) \equiv \text{Leading order in QCDf} + h_\lambda(q^2)$, with $h_\lambda$ denoting the unknown power corrections. 
The short-distance NP contributions due to $C_9$ (and/or $C_7$) can be mimicked by long-distance effects in $h_\lambda$. Hence, the estimation of the size of the hadronic contributions is crucial in determining whether the observed deviations are due to new physics. 

It is possible to parameterise the power corrections by a polynomial with a number of free parameters to be fitted to the experimental data~\cite{Jager:2014rwa}. Assuming a $q^2$-polynomial ansatz~\cite{Hurth:2020rzx,Chobanova:2017ghn} we can write 
\begin{align}
  h_\pm(q^2)&= h_\pm^{(0)} + \frac{q^2}{1 \,{\rm GeV}^2}h_\pm^{(1)} + \frac{q^4}{1 \,{\rm GeV}^4}h_\pm^{(2)}\,,\\
h_0(q^2)&= \sqrt{q^2}\times \left( h_0^{(0)} + \frac{q^2}{1\, {\rm GeV}^2}h_0^{(1)} + \frac{q^4}{1\, {\rm GeV}^4}h_0^{(2)}\right)\,,\nonumber
\end{align}
which is the most general ansatz for the unknown hadronic contributions (up to higher order powers in $q^2$) compatible with the analyticity structure assumed in Ref.~\cite{Bobeth:2017vxj}. 

By doing separate fits for NP and hadronic parameters, a statistical comparison between the two fits is possible using the Wilks' theorem~\cite{Wilks:1938dza}. As was shown in Ref.~\cite{Arbey:2018ics}, while NP explanation seems to be favoured, more data is needed to be able to make conclusive statements.

\section{Global fits}
Global statistical fits of the Wilson coefficients with the available data are a standard way to search for evidence for NP contributions, in a way that is agnostic to the specific NP model.

In order to check the coherence of the implication of the clean observables for new physics compared to the rest of the observables, we first present a fit to the former where we consider BR($B_{s,d} \to \mu^+ \mu^-$), $R_K, R_{K^*}$ as well as the recently measured $R_{K_S^0}$ and $R_{K^{*+}}$~\cite{LHCb:2021lvy}.   
The calculation of the observables and the $\chi^2$ fitting is done with the SuperIso public program~\cite{Mahmoudi:2007vz,Mahmoudi:2008tp,Mahmoudi:2009zz,Neshatpour:2021nbn,Neshatpour:2022fak}.

\subsection{Clean observables}
In Table~\ref{tab:1D_Full2021_clean} the one-dimensional NP fit to clean observables are provided.
Compared to the NP fit to the clean observables in Ref.~\cite{Hurth:2021nsi}, the two recently measured LFUV ratios $R_{K_S^0}$ and $R_{K^{*+}}$~\cite{LHCb:2021lvy} as well as the $R_K$ measurement by Belle~\cite{BELLE:2019xld} in the [1,6] GeV$^2$ bin are included in the fit. 
\begin{table}[t]
\begin{center}
\caption{Fit to clean observables with the full data from 2021.}
\setlength\extrarowheight{5pt}
\scalebox{1}{
\begin{tabular}{|C{1.2cm}|C{2.7cm}|C{1.7cm}|C{1.7cm}|}
\hline 
 \multicolumn{4}{|c|}{\footnotesize Only LFUV ratios and $B_{s,d}\to \ell^+ \ell^-$ \vspace{-0.1cm}} \\
 \multicolumn{4}{|c|}{($\chi^2_{\rm SM}=34.25,\;$ nr. obs.$=12$)} \\ \hline
    & b.f. value & $\chi^2_{\rm min}$ & ${\rm Pull}_{\rm SM}$  \\ 
\hline \hline
$\delta C_{9} $    	& $ 	-2.00	\pm	5.00	 $ & $ 	34.1	 $ & $	0.4	\sigma	 $  \\
$\delta C_{9}^{e} $    	& $ 	0.83	\pm	0.21	 $ & $ 	14.5	 $ & $	4.4	\sigma	 $  \\
$\delta C_{9}^{\mu} $    	& $ 	-0.80	\pm	0.21	 $ & $ 	15.4	 $ & $	4.3	\sigma	 $  \\
\hline										
$\delta C_{10} $    	& $ 	0.43	\pm	0.24	 $ & $ 	30.6	 $ & $	1.9	\sigma	 $  \\
$\delta C_{10}^{e} $    	& $ 	-0.81	\pm	0.19	 $ & $ 	12.3	 $ & $	4.7	\sigma	 $  \\
$\delta C_{10}^{\mu} $    	& $ 	0.66	\pm	0.15	 $ & $ 	10.3	 $ & $	4.9	\sigma	 $  \\
\hline							          			
\hline
$\delta C_{\rm LL}^e$	& $ 	0.43	\pm	0.11	 $ & $ 	13.3	 $ & $	4.6	\sigma	 $  \\
$\delta C_{\rm LL}^\mu$    	& $ 	-0.39	\pm	0.08	 $ & $ 	10.1	 $ & $	4.9	\sigma	 $  \\
\hline										
\end{tabular}
} 
\label{tab:1D_Full2021_clean} 
\end{center} 
\end{table}
\begin{figure}[h]
\centering
\includegraphics[width=80mm]{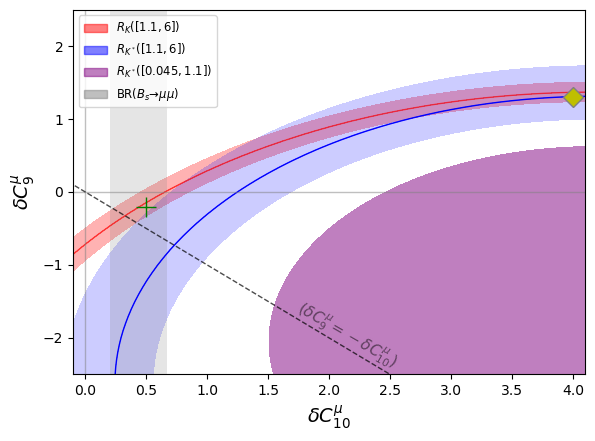}\\
\caption{The coloured and grey regions correspond to the values of $\delta C_9^\mu$ and $\delta C_{10}^\mu$ that result in a prediction of $R_{K^{(*)}}$ and BR($B_s\to\mu^+\mu^-$) within $1\sigma \left(=\sqrt{\sigma_{\rm th}^2 + \sigma_{\rm exp}^2}\right)$ of their measured values, respectively. The solid red (blue) line corresponds to the LHCb measured central value for $R_{K^{(*)}}$ in the $q^2\in[1.1,6]$ GeV$^2$ bin.
The yellow diamond (green cross) indicates the best fit value to  $R_{K^{(*)}}$  excluding (including) BR($B_s\to\mu^+\mu^-$). See caption of Fig.~3 in Ref.~\cite{Hurth:2021nsi} for further information.} 
\label{fig:BsmumuRole}
\end{figure}
While the hierarchy of the most favoured scenarios has remained the same, the addition of the two former measurements has resulted in an increase in the NP significance compared to~\cite{Hurth:2021nsi} ($\sim0.4\sigma$ for NP in $C_{\rm LL}^{e,\mu}$, $C_{10}^{e,\mu}$ and $~0.3\sigma$ for $C_{9}^{e,\mu}$).
The inclusion of BR($B_s \to \mu^+ \mu^-$) in this set of observables is crucial in breaking the degeneracy between NP in $C_9^\mu$ and $C_{10}^\mu$ for explaining the measured values of the ratios (see Fig.~\ref{fig:BsmumuRole}).

\begin{table}[t]
\begin{center}
\caption{Fit to all except the clean observables with the full data from 2021.}
\setlength\extrarowheight{5pt}
\scalebox{1}{
\begin{tabular}{|C{1.2cm}|C{2.7cm}|C{1.7cm}|C{1.7cm}|}
\hline 
 \multicolumn{4}{|c|}{\footnotesize All observables except LFUV ratios and $B_{s,d}\to \ell^+ \ell^-$ \vspace{-0.1cm}} \\ 
 \multicolumn{4}{|c|}{($\chi^2_{\rm SM}=221.8,\;$ nr. obs.$=171$)} \\ \hline
                          & b.f. value & $\chi^2_{\rm min}$ & ${\rm Pull}_{\rm SM}$  \\ 
\hline \hline
$\delta C_{9} $    	& $ 	-0.95	\pm	0.13	 $ & $ 	185.1	 $ & $	6.1	\sigma	 $  \\
$\delta C_{9}^{e} $    	& $ 	0.70	\pm	0.60	 $ & $ 	220.5	 $ & $	1.1	\sigma	 $  \\
$\delta C_{9}^{\mu} $    	& $ 	-0.96	\pm	0.13	 $ & $ 	182.8	 $ & $	6.2	\sigma	 $  \\
\hline										
$\delta C_{10} $    	& $ 	0.29	\pm	0.21	 $ & $ 	219.8	 $ & $	1.4	\sigma	 $  \\
$\delta C_{10}^{e} $    	& $ 	-0.60	\pm	0.50	 $ & $ 	220.6	 $ & $	1.1	\sigma	 $  \\
$\delta C_{10}^{\mu} $    	& $ 	0.35	\pm	0.20	 $ & $ 	218.7	 $ & $	1.8	\sigma	 $  \\
\hline							          			
\hline
$\delta C_{\rm LL}^e$	& $ 	0.34	\pm	0.29	 $ & $ 	220.6	 $ & $	1.1	\sigma	 $  \\
$\delta C_{\rm LL}^\mu$    	& $ 	-0.64	\pm	0.13	 $ & $ 	195.0	 $ & $	5.2	\sigma	 $  \\
\hline										
\end{tabular}
} 
\label{tab:1D_Full2021_rest} 
\end{center} 
\end{table}
\subsection{All observables except the clean ones}\label{sec:rest}
In Table~\ref{tab:1D_Full2021_rest} we give the one-dimensional NP fit to all observables except the LFUV ratios and $B_{s,d}\to \ell^+ \ell^-$. 
In this set, the significance of the NP fit is dependent on the assumption for the size of the non-factorisable power corrections. The results of Table~\ref{tab:1D_Full2021_rest} is given with the assumption of 10\% power correction.
Compared to Ref.~\cite{Hurth:2021nsi}, the previous LHCb results of the $B_s \to \phi \mu^+\mu^-$ observables are replaced by the recent measurements with an integrated luminosity of 8.4 fb$^{-1}$~\cite{LHCb:2021xxq,LHCb:2021zwz}. We also include the $F_H(B^+\to K^+\mu^+\mu^-)$ measurement by CMS~\cite{CMS:2018qih} as well as the $B\to K^* e^+ e^-$ angular observable measurements by LHCb~\cite{LHCb:2020dof}.
With these updates, the hierarchy of the most favoured scenarios has remained the same, with the largest significance for NP in vector lepton current in the muon sector $\delta C_9^\mu$ or by lepton flavour universal NP in $\delta C_9$ followed by NP in the chiral basis $\delta C_{\rm LL}^\mu$ in the muon sector.
These three favoured scenarios are all now showing a reduction of $\sim 0.4\sigma$ compared to Ref.~\cite{Hurth:2021nsi} which can be attributed mostly to the updated measurement of the $B_s \to \phi \mu^+ \mu^-$ observables.

The fit with all observables except the clean ones shows no indication for NP in the electron sector, however, besides the agreement of the measurements in the electron sector with their SM predictions, it should be noted that in this set of observables the experimental data in the electron mode is much scarcer compared to the muon sector. 
Comparison of Tables~\ref{tab:1D_Full2021_clean} and~\ref{tab:1D_Full2021_rest} shows that there is not a complete consistency between all favoured scenarios for each of the two datasets. However, there are scenarios such as NP in $\delta C_9^\mu$ for which not only there is a large significance for both datasets but also the best fit values agree within $1\sigma$.

\subsection{Global fit to all \texorpdfstring{$b \to s \ell^+ \ell^-$}{b->sll} observables}
To get the global picture of the rare $B$-decays, we consider here all the relevant observables (consisting of the clean and the rest of the observables mentioned in the previous subsections). For the global fit, similar to the fits of subsection~\ref{sec:rest}, we consider 10\% uncertainty for the power corrections.

\subsubsection{One-dimensional fit}

\begin{table}[b!]
\begin{center}
\caption{Fit to all observables with the full data from 2021.}
\setlength\extrarowheight{5pt}
\scalebox{1}{
\begin{tabular}{|C{1.2cm}|C{2.5cm}|C{1.7cm}|C{1.7cm}|}
\hline 
\multicolumn{4}{|c|}{All observables} \\[-4pt]										
\multicolumn{4}{|c|}{($\chi^2_{\rm SM}=253.3,\;$ nr. obs.$=183$)} \\ \hline			
& b.f. value & $\chi^2_{\rm min}$ & ${\rm Pull}_{\rm SM}$  \\										
\hline \hline										
$\delta C_{9} $    	& $ 	-0.93	\pm	0.13	 $ & $ 	218.4	 $ & $	5.9	\sigma	 $  \\
$\delta C_{9}^{e} $    	& $ 	0.82	\pm	0.19	 $ & $ 	232.3	 $ & $	4.6	\sigma	 $  \\
$\delta C_{9}^{\mu} $    	& $ 	-0.90	\pm	0.11	 $ & $ 	197.7	 $ & $	7.5	\sigma	 $  \\
\hline										
$\delta C_{10} $    	& $ 	0.27	\pm	0.17	 $ & $ 	250.5	 $ & $	1.7	\sigma	 $  \\
$\delta C_{10}^{e} $    	& $ 	-0.78	\pm	0.18	 $ & $ 	230.4	 $ & $	4.8	\sigma	 $  \\
$\delta C_{10}^{\mu} $    	& $ 	0.54	\pm	0.12	 $ & $ 	231.5	 $ & $	4.7	\sigma	 $  \\
\hline							          			
\hline										
$\delta C_{\rm LL}^e$	& $ 	0.42	\pm	0.10	 $ & $ 	231.2	 $ & $	4.7	\sigma	 $  \\
$\delta C_{\rm LL}^\mu$    	& $ 	-0.46	\pm	0.07	 $ & $ 	208.2	 $ & $	6.7	\sigma	 $  \\
\hline										
\end{tabular}
} 
\label{tab:1D_full2021_all} 
\end{center} 
\end{table}
In Table~\ref{tab:1D_full2021_all} the one-dimensional global fits of NP to all relevant rare $B$-decays are given. As expected, the scenario with highest significance corresponds to NP in $\delta C_9^\mu$, this is followed by NP in $\delta C_{\rm LL}^\mu$ and the lepton flavour universal contribution $\delta C_9$. This pattern is similar to what was observed in Ref.~\cite{Hurth:2021nsi}, while now the significance has reduced very slightly ($\sim 0.1\sigma$) for $\delta C_9^\mu$ and $\delta C_{\rm LL}^\mu$ and $\sim 0.4\sigma$ for $\delta C_9$. In Table~\ref{tab:1D_full2021_all} we have not considered scenarios with NP contribution  to the electromagnetic dipole coefficient $C_7$ or the (pseudo)scalar coefficients ($C_{Q_{1,2}}$) since in a one-dimensional fit these get highly constrained, the former by $B\to X_s \gamma$ and the latter by $B_s\to\mu^+\mu^-$. We have also not given NP fits to contribution to right-handed quark currents ($\delta C_i^{\prime}$) which are not favoured by the data.
However in general in a multidimensional fit the situation can change as for example pointed out for the pseudo(scalar) contributions which when varied together with $C_{10}$ can admit a large range of values~\cite{Arbey:2018ics}.

\subsubsection{Multidimensional fit}
\begin{table}[b!]
\begin{center}
\caption{20-dimensional fit to all observables with the full data from 2021.}
\setlength\extrarowheight{5pt}
\scalebox{1}{
\begin{tabular}{|C{1.9cm}|C{1.9cm}|C{1.9cm}|C{1.9cm}|}
\hline																
\multicolumn{4}{|c|}{All observables  with $\chi^2_{\rm SM}=253.3,\;$ nr. obs.$=183$} \\											
\multicolumn{4}{|c|}{($\chi^2_{\rm min}=	 	179.2	;\; {\rm Pull}_{\rm SM}=	5.5	(	5.5	)	\sigma$)} \\								
\hline \hline																
\multicolumn{2}{|c|}{$\delta C_7$} &  \multicolumn{2}{c|}{$\delta C_8$}\\																
\multicolumn{2}{|c|}{$	0.06	\pm	0.03	$} & \multicolumn{2}{c|}{$	-0.80	\pm	0.40	$}\\								
\hline																
\multicolumn{2}{|c|}{$\delta C_7^\prime$} &  \multicolumn{2}{c|}{$\delta C_8^\prime$}\\																
\multicolumn{2}{|c|}{$	-0.01	\pm	0.01	$} & \multicolumn{2}{c|}{$	-0.30	\pm	1.30	$}\\								
\hline																
$\delta C_{9}^{\mu}$ & $\delta C_{9}^{e}$ & $\delta C_{10}^{\mu}$ & $\delta C_{10}^{e}$ \\																
$	-1.15	\pm	0.18	$ & $	-6.60	\pm	1.60	$ & $	0.21	\pm	0.20	$ & $	2.60	\pm	2.7	$ \\
\hline\hline																
$\delta C_{9}^{\prime \mu}$ & $\delta C_{9}^{\prime e}$ & $\delta C_{10}^{\prime \mu}$ & $\delta C_{10}^{\prime e}$ \\																
$	0.05	\pm	0.31	$ & $	1.40	\pm	2.10	$ & $	-0.04	\pm	0.19	$ & $	1.30	\pm	2.8	$ \\
\hline\hline																
$C_{Q_{1}}^{\mu}$ & $C_{Q_{1}}^{e}$ & $C_{Q_{2}}^{\mu}$ & $C_{Q_{2}}^{e}$ \\																
$	0.07	\pm	0.06	$ & $	-1.60	\pm	1.60	$ & $	-0.11	\pm	0.14	$ & $	-4.00	\pm	1.2	$ \\
\hline\hline																
$C_{Q_{1}}^{\prime \mu}$ & $C_{Q_{1}}^{\prime e}$ & $C_{Q_{2}}^{\prime \mu}$ & $C_{Q_{2}}^{\prime e}$ \\																
$	-0.07	\pm	0.06	$ & $	-1.70	\pm	1.30	$ & $	-0.21	\pm	0.15	$ & $	-4.10	\pm	0.8	$ \\
\hline																
\end{tabular}
} 
\label{tab:20D_full2021_all} 
\end{center} 
\end{table}
In general, UV-complete new physics models can contribute to several Wilson coefficients in the Weak effective theory, it is thus reasonable to make a multidimensional fit where more than one Wilson coefficient can be varied. Table~\ref{tab:20D_full2021_all} presents the result of a twenty-dimensional fit to the rare $B$ decay observables when varying all Wilson coefficients. Within this approach we avoid any look elsewhere effect (LEE).
In general, LEE gets introduced when concentrating on a subset of observables, it also takes place if one and/or two specific NP directions are assumed a posteriori. However, this is avoided when assuming all possible Wilson coefficients.
In principle, there might be insensitive coefficients and flat directions leading to an underestimation of the significance of the multidimensional fit. However, by considering likelihood profiles and correlation matrices we eliminate these coefficients, finding an ``effective'' number of degrees of freedom (for the 20-dimensional fit we find 19 effective dof). 
The significance of the 20-dimensional fit is $5.5\sigma$, similar to what was obtained in Ref.~\cite{Hurth:2021nsi}.
Several of the Wilson coefficients are still only loosely constrained as less data with electrons in the final state is available than with muons in the final state.

\section{Comparison of global fit results}
\begin{figure}[t!]
\centering
\includegraphics[width=70mm]{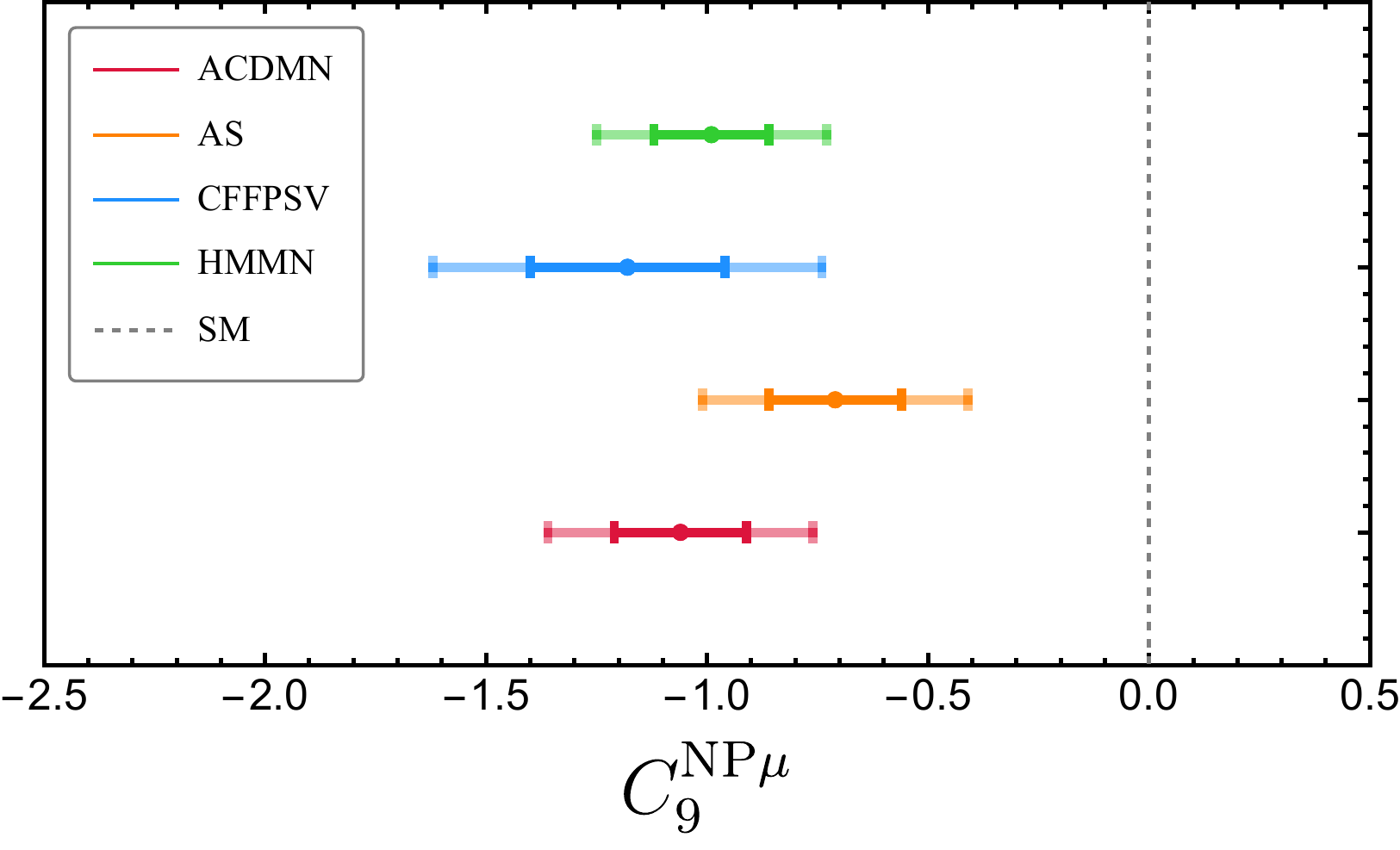}\\
\includegraphics[width=67mm]{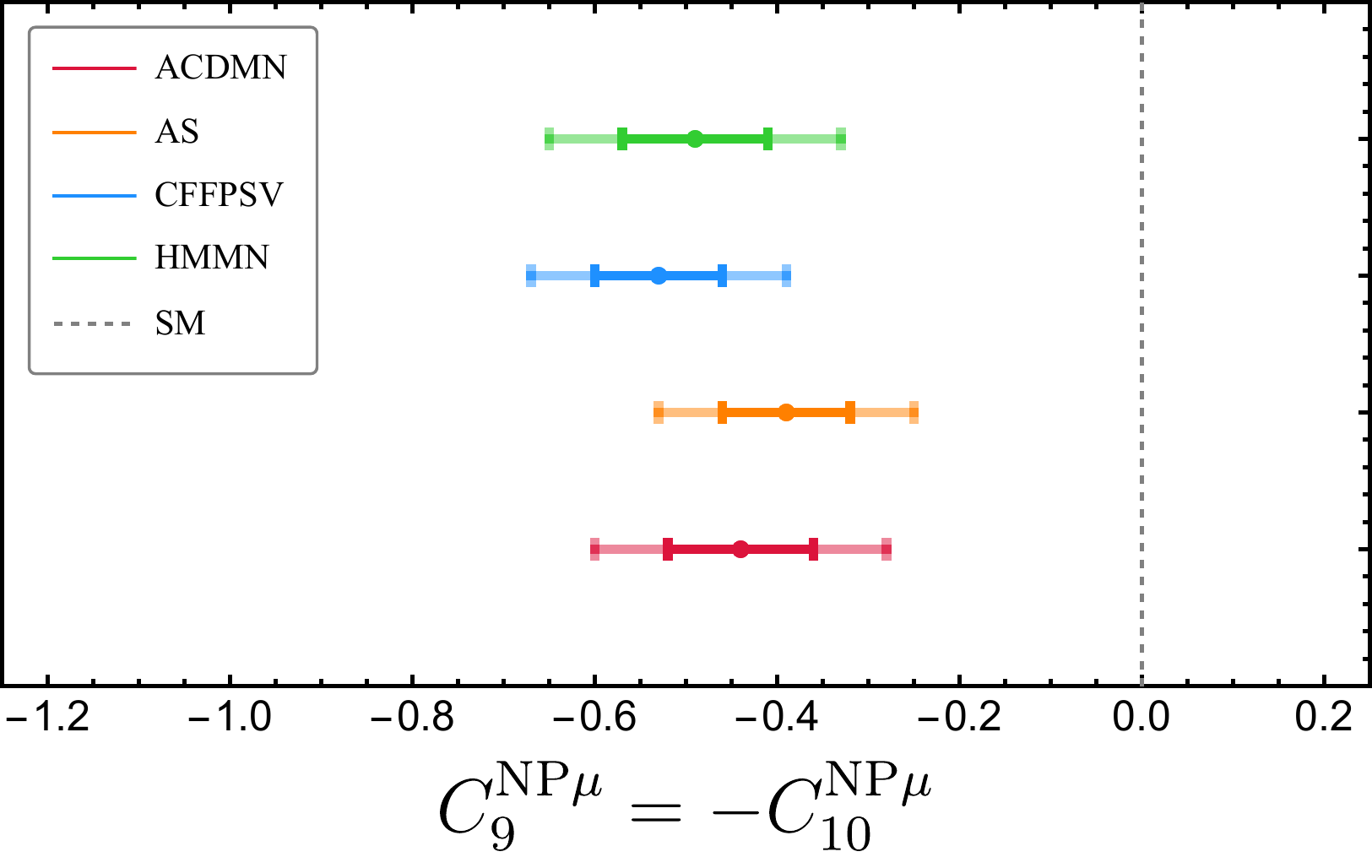}\\
\caption{Comparison of one-dimensional fits for NP contributions to $C_9^\mu$ (top) and to $C_9^\mu = -C_{10}^\mu$ (bottom). The darker (lighter) coloured lines correspond to 68\% (95\%) CL intervals. See the main text for the definition and the relevant reference for each group.} 
\label{fig:comp_1d}
\end{figure}

\begin{figure}[t!]
\centering
\includegraphics[width=80mm]{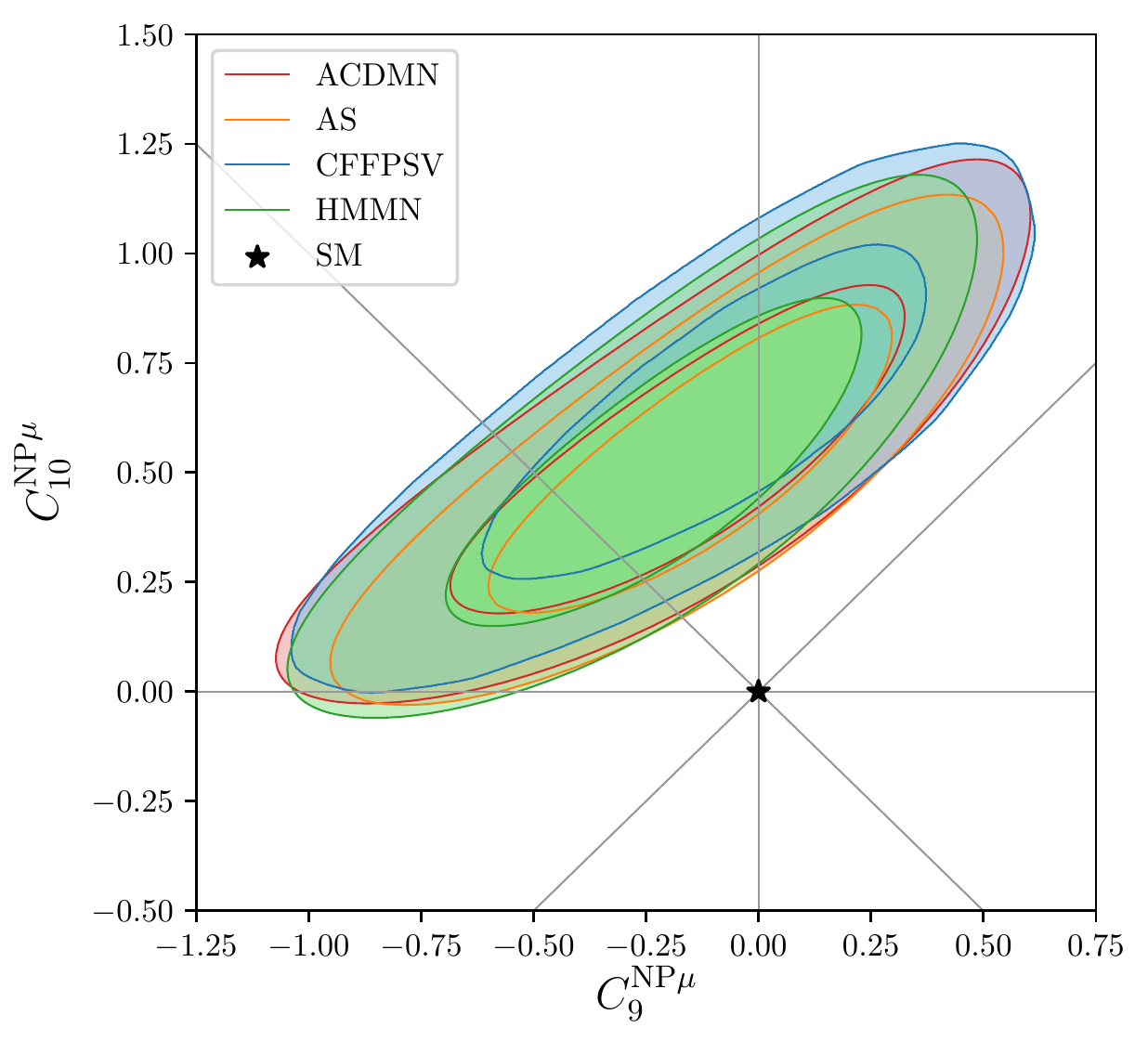}\\
\includegraphics[width=80mm]{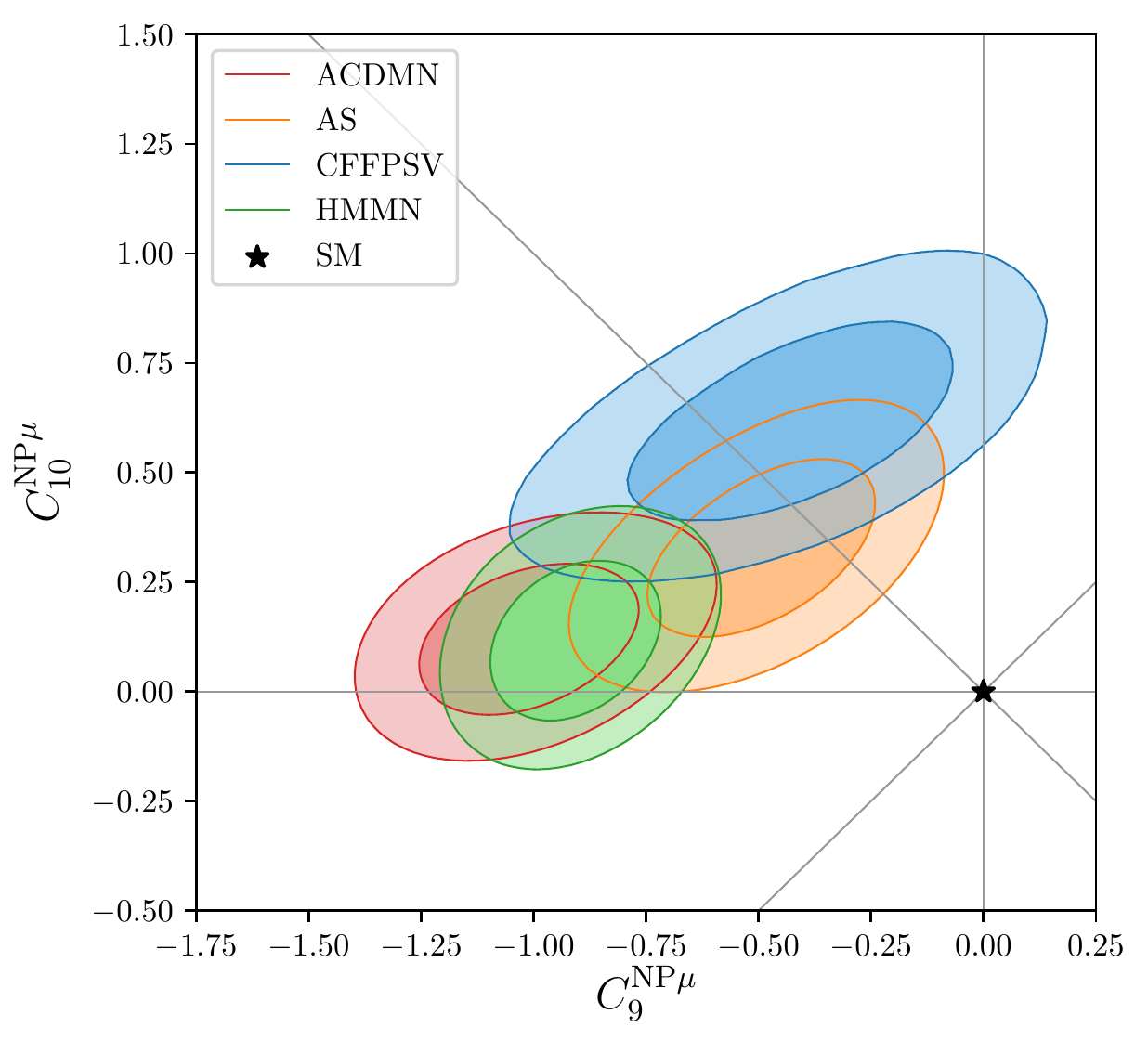}\\
\caption{Comparison of two-dimensional fits for NP contributions to \{$C_9^\mu, C_{10}^\mu$\} using the clean observables (top) and using all available $b\to s \ell\ell$ data (bottom). The darker (lighter) coloured regions correspond to 68\% (95\%) CL fit results. See the main text for the definition and relevant reference for each group.} 
\label{fig:comp_2d}
\end{figure}

Here we present a comparison between global fits performed by the different fitting groups, based on the common work presented at the Flavour Anomaly Workshops~\cite{durham1,durham}, where the results of the following groups have been confronted:\\[0.2cm]
- ACDMN: M. Algueró, B. Capdevila, S. Descotes-Genon, J. Matias, M. Novoa-Brunet~\cite{Alguero:2021anc}.\\[0.2cm]
- AS: W. Altmannshofer, P. Stangl~\cite{Altmannshofer:2021qrr}.\\[0.2cm]
- CFFPSV: M. Ciuchini, M. Fedele, E. Franco, A. Paul, L. Silvestrini, M. Valli~\cite{Ciuchini:2020gvn}.\\[0.2cm]
- HMMN: T. Hurth, F. Mahmoudi, D. Martínez-Santos, S. Neshatpour~\cite{Hurth:2021nsi}.\\[0.2cm]
Similar fits have also been performed in \cite{Geng:2021nhg,Bhom:2020lmk,Alok:2019ufo,Datta:2019zca,Kowalska:2019ley}.
Fig.~\ref{fig:comp_1d} shows the one-dimensional global fit results for $C_9^\mu$ and $C_9^\mu = -C_{10}^\mu$, and Fig.~\ref{fig:comp_2d} shows the two-dimensional fits for clean observables (upper plot) and for all $bs\ell\ell$ observables (lower plot).

While there are differences in experimental inputs, form factors, assumptions on non-local matrix elements and statistical frameworks considered by different groups (see~\cite{durham1,durham} for more details), Figs.~\ref{fig:comp_1d} and \ref{fig:comp_2d} show a remarkable global agreement between different results and the robustness of the conclusions.

%

\section{Projections for clean observables}
\begin{figure}[h!]
\centering
\includegraphics[width=80mm]{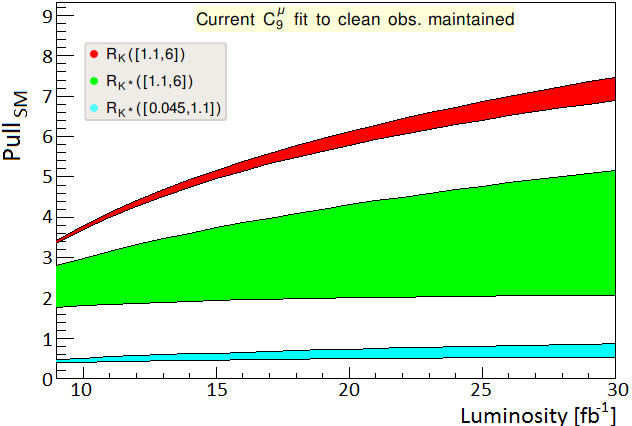}\\
\includegraphics[width=80mm]{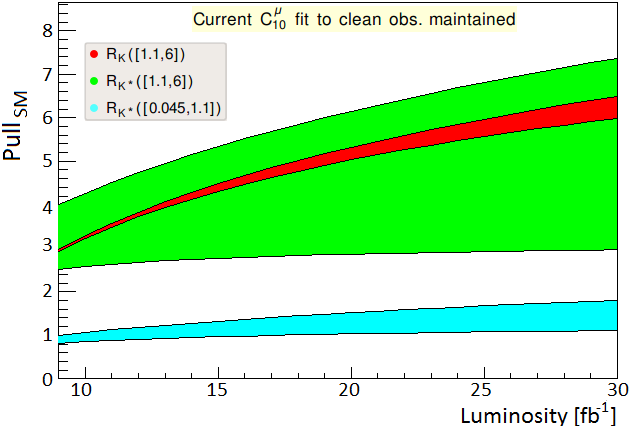}\\
\caption{Pull$_{\rm SM}$ for $R_K[1.1,6]$, $R_{K^*}[1.1,6]$ and $R_{K^*}[0.045,1.1]$ as a function of LHCb integrated luminosity assuming the current best fit point for NP in $C_9^\mu$ ($C_{10}^\mu$) for the top (bottom) plot.} 
\label{fig:clean_proj_individual}
\end{figure}
The global fit to rare $B$-decay observables suggests several NP scenarios with quite large significances. However, these significances are dependent on the assumed size of non-factorisable power corrections.  It is thus useful to check when the clean observables (which are theoretically very precisely predicted) can individually reach a $5\sigma$ significance. In Fig.~\ref{fig:clean_proj_individual}, assuming the current best fit value from the clean observables in the one-dimensional fit to $\delta C_9^\mu$ ($\delta C_{10}^\mu$) remains, in the upper (lower) plot we have shown the significance of $R_K$ and $R_{K^{*}}$ as a function of the luminosity with the upper and lower bound for each observable corresponding to two different assumptions on the systematic uncertainties~\cite{Aaij:2244311} (for further details see Ref.~\cite{Hurth:2021nsi}). From Fig.~\ref{fig:clean_proj_individual} it is clear that if the current fit to $\delta C_9^\mu$ or to  $\delta C_{10}^\mu$ is the correct description of new physics, $R_K$ can reach $5\sigma$ already with less than 20 fb$^{-1}$ of data.

\begin{figure}[h] 
\centering
\includegraphics[width=80mm]{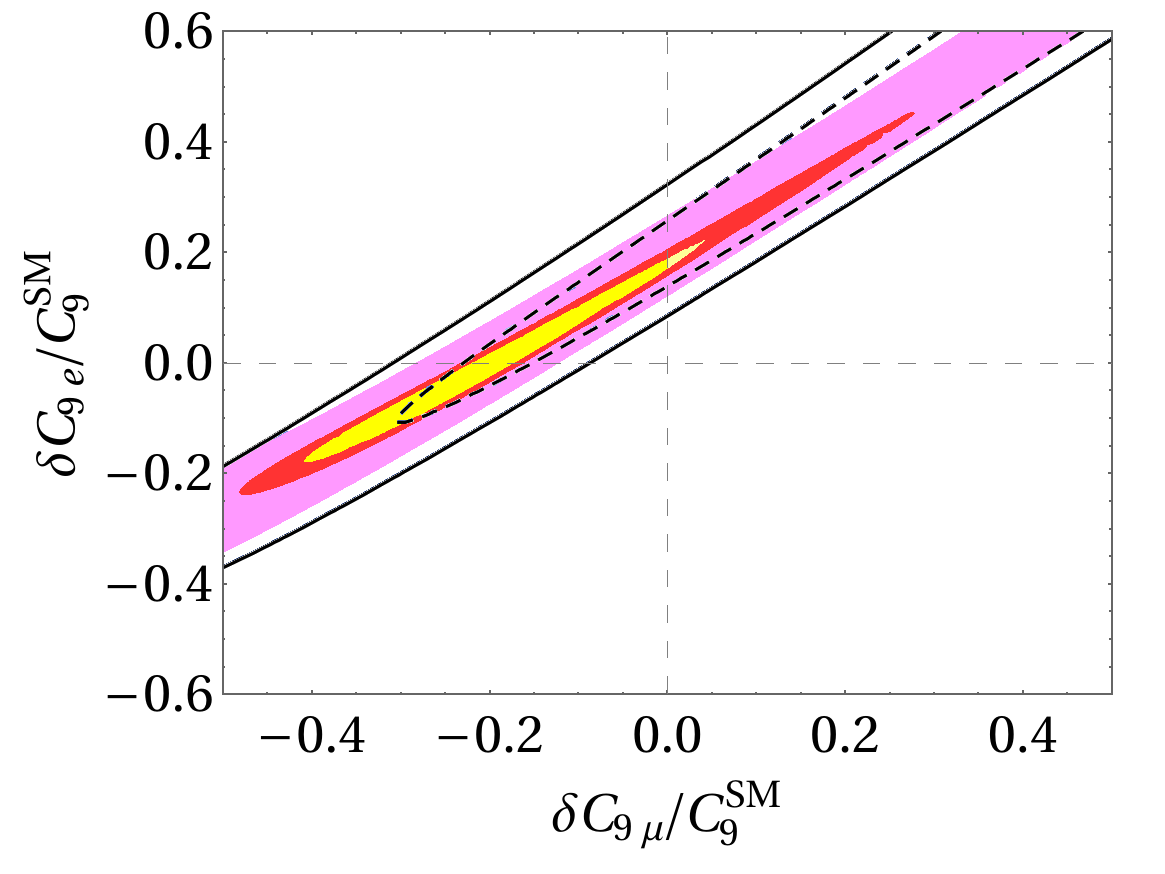}\\
\includegraphics[width=80mm]{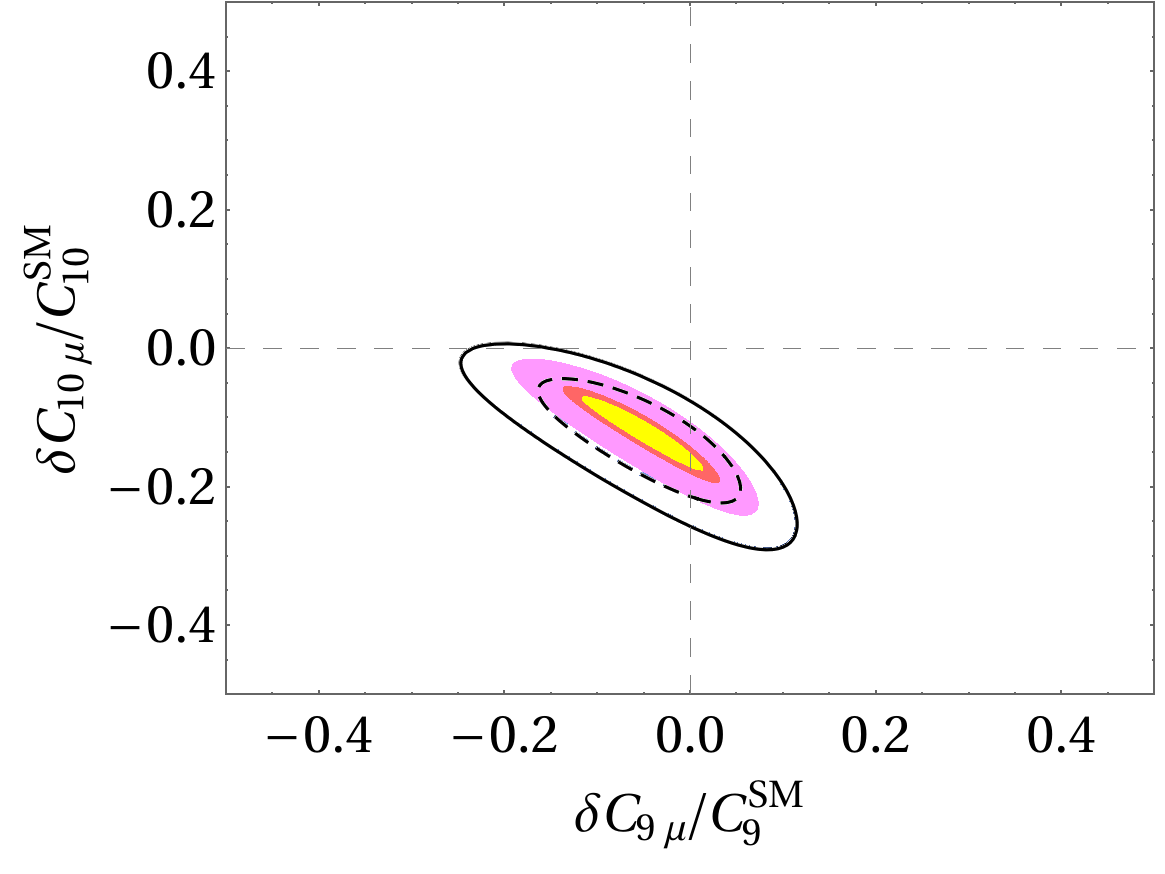}\\
\caption{Projection of the clean observables, assuming current $\{C_{9}^\mu, C_{9}^e\}$ ($\{C_{9}^\mu, C_{10}^\mu\}$) two-dimensional best fit point remains in the top (bottom) plot.
The dashed and solid lines indicate the contour of the 68\% and 95\% CL fit with the current data, respectively. The pink, red and yellow regions correspond to 95\% CL projections with 18, 50 and 300 fb$^{-1}$ of data, respectively.} 
\label{fig:clean_proj}
\end{figure}

We also make projections for the two-dimensional fits to all clean observables assuming the current best fit values for $\{C_9^e,C_{9}^\mu\}$ or $\{C_9^\mu,C_{10}^\mu\}$ remain. The results are given in Fig.~\ref{fig:clean_proj} for the planned LHCb upgrades, for three benchmark points with 18, 50 and 300 fb$^{-1}$ integrated luminosities. The best fit points for each scenario has a Pull$_{\rm SM}$ larger than $5\sigma$ already with 18 fb$^{-1}$ data.

\section{Conclusions}
The updated NP fits to rare $B$ decays including the updated measurement of $B_s \to \phi \mu^+ \mu^-$ observables as well as the recent measurement of lepton flavour violating ratios $R_{K^{*+}}$ and $R_{K_S}$ by LHCb follow the same trend as with the previous set of results favouring in particular new physics contributions in the Wilson coefficient $C_9^\mu$, with an increased significance. Interestingly, while the updated measurements have slightly changed the NP significance, the hierarchy of the preferred scenarios has remained the same. The projections for clean observables, show that if the current tensions remain, $R_K$ can establish NP with $5\sigma$ significance already with less than 20 fb$^{-1}$ of data. The main source of theory uncertainty in global fits is due to non-local hadronic contributions. However, different fits with different setups, inputs and statistical frameworks show a remarkable agreement so that the experimental observation of the discrepancy in these observables would be a clear sign of physics beyond the SM.

\begin{acknowledgments}
FM would like to thank the organisers of the FPCP 2022 conference for their invitation to present this talk, and for the very fruitful conference. She is also grateful to her collaborators and in particular to S. Neshatpour for his help in updating the fit results. 

\end{acknowledgments}

\bigskip 

\bibliography{fpcp2022}

\end{document}